\documentclass{article}
\usepackage[T1]{fontenc} 
\usepackage[utf8]{inputenc} 
\usepackage{ismir,amsmath,cite,url,amssymb,verbatim}
\usepackage{graphicx}
\usepackage{color}

\usepackage{lineno}

\title{S\lowercase{pec}TNT: a Time-Frequency Transformer for Music Audio}

\multauthor
{Wei-Tsung Lu$^1$ \hspace{0.5cm} Ju-Chiang Wang$^1$ \hspace{0.5cm} Minz Won$^{1,2}$ \hspace{0.5cm} Keunwoo Choi$^1$ \hspace{0.5cm} Xuchen Song$^1$} 
{
	$^1$ ByteDance, Mountain View, California, United States\\
	$^2$ Music Technology Group, Universitat Pompeu Fabra, Barcelona, Spain\\
{\tt\small \{weitsung.lu, ju-chiang.wang, minzwon, keunwoo.choi, xuchen.song\}@bytedance.com}
}

\def\authorname{W.-T. Lu, J.-C. Wang, M. Won, K. Choi, and X. Song}

\begin{document}

\maketitle
\begin{abstract}

Transformers have drawn attention in the MIR field for their remarkable performance shown in natural language processing and computer vision. However, prior works in the audio processing domain mostly use Transformer as a temporal feature aggregator that acts similar to RNNs. In this paper, we propose SpecTNT, a Transformer-based architecture to model both spectral and temporal sequences of an input time-frequency representation. Specifically, we introduce a novel variant of the Transformer-in-Transformer (TNT) architecture. In each SpecTNT block, a spectral Transformer extracts frequency-related features into the frequency class token (FCT) for each frame. Later, the FCTs are linearly projected and added to the temporal embeddings (TEs), which aggregate useful information from the FCTs. Then, a temporal Transformer processes the TEs to exchange information across the time axis. By stacking the SpecTNT blocks, we build the SpecTNT model to learn the representation for music signals. In experiments, SpecTNT demonstrates state-of-the-art performance in music tagging and vocal melody extraction, and shows competitive performance for chord recognition. The effectiveness of SpecTNT and other design choices are further examined through ablation studies.  

\end{abstract}

\section{Introduction}\label{sec:introduction}
Deep learning models have been actively used in recent music information retrieval (MIR) research. Although the spirit of deep learning is end-to-end learning, however, various assumptions are made during making design choices of deep learning models. 

Regarding assumptions on spectrograms, the most popular form of music audio representation in deep learning, the time-axis is often considered to be the axis of \textit{sequence} while the frequency-axis is the axis of \textit{feature}. For example, in \cite{rigaud2016singing, leglaive2015singing}, recurrent layers were applied to model a spectrogram as a sequence of spectra. In \cite{dieleman2014end}, convolutional layers were used to aggregate features over time after the first convolutional layer models multiple frames of spectra as feature. On the other hand, in \cite{choi2016automatic}, two-dimensional convolutional layers were used, equating the frequency- and time-axes. There are also hybrid approaches, such as convolutional recurrent neural networks (CRNN)~\cite{choi2017convolutional} and convolutional Transformer \cite{won2019toward}, in which recurrent layers or Transformer are applied along the time axis.

In spectrograms, it is well known that there are meaningful spectral patterns. Different music components exist in different frequency ranges, and there is a very strong spectral correlation called harmonics. Since a normal convolutional layer can model local patterns only, several approaches have been proposed to model harmonics along the frequency axis. Harmonic Constant-Q Transform (HCQT) is a novel multi-channel time-frequency representation that was proposed to overcome the limitation by improving the input representation of audio \cite{bittner2017deep}. Harmonic CNNs (convolutional neural networks) \cite{won2020data} are designed to model the harmonic pattern by modifying the convolutional filters. However, these solutions only model some of the spectral patterns, reminding the need for a more general solution with higher flexibility. 


Transformers have successfully demonstrated their ability to model the sequential data with long-term (inter-) dependency and invariance. This is achieved by multiple aspects of Transformers. First, the key-query mechanism enables modeling the relationship of every combination of the instances. Second, positional encoding helps the model to take the order of instances into account. Through stacked attention layers, the input sequence is transformed into a sequence of representations that are based on the inter-dependency of the input.
The prior works in audio analysis are mostly based on a similar, naive approach where Transformer is used as a temporal feature aggregator that acts similar to RNNs (recurrent neural networks), with few exceptions such as \cite{zadeh2019wildmix}. 

Recently, Transformer in Transformer (TNT), a variant of Transformer that arranges two Transformers in a hierarchical manner, was proposed~\cite{han2021transformer} for image recognition. 
In TNT, an inner (lower-level) Transformer is applied to extract the local pixel-level embeddings, and then the pixel-level embeddings are projected to the patch-level embedding space which is later handled by an outer (higher-level) Transformer to summarize a global representation. One can simply apply TNT for audio by treating a song as an image, which is comprised of a sequence of frames (patches) while the frequency bins within frames are considered as pixels. However, from our pilot study, we find this approach results in unstable training and only achieves similar performance compared to using the original Transformer. This is possibly because the amount of training data is insufficient or the interpretation of frequency sequences is different from that of pixel sequences. Therefore, non-trivial modifications from the original idea of TNT should be made. 

In this paper, we propose \emph{SpecTNT}, a time-frequency transformer that models spectrograms as a sequence along both time- and frequency-axes. Similar to TNT, SpecTNT uses two Transformers hierarchically. However, the temporal local embeddings extracted from the inner Transformer are not directly sent to the outer Transformer. Instead, a special token called \emph{frequency class token} (FCT) is appended to aggregate the important spectral features of each frame. The FCT is then projected to the global (temporal) embedding space to enable the information exchange across the time axis. This design allows the important local information is passed to the outer Transformer through FCT while reducing the dimensionality of the data flow compared to the original TNT. As a result, it helps SpecTNTs to perform well on audio-related tasks even with smaller datasets.

Our contributions can be summarized as follows: 
(1) to the best of our knowledge, our work is the first attempt to leverage TNT-based architecture to learn the representations for audio; (2) we propose SpecTNT, a novel modification of TNT to better fit the music data for MIR tasks; (3) we conduct extensive experiments to demonstrate the capability of SpecTNT in various MIR tasks -- vocal melody extraction, music auto-tagging, and chord recognition.


\section{Related Work}

In this section, we review the literature in music tagging, vocal melody extraction, and chord recognition -- three well-defined MIR tasks adopted in the experiments to evaluate SpecTNT.
Due to space limitation, we focus on the recent trends since the adoption of deep learning approaches. 

Music tagging is a multi-label classification task that annotates a music audio clip with various types of labels such as genres (rock, jazz), instruments (vocal, guitar, drums), and mood (happy, sad)~\cite{lamere2008social}. Since a CNN-based approach has been first introduced~\cite{dieleman2014end}, various advanced architectures have been used including a two-dimensional CNN~\cite{choi2016automatic}, a sample-level CNN~\cite{lee2017sample}, and a two-dimensional ResNet~\cite{won2020evaluation}. Due to the open nature of the tag set, among MIR tasks, music tagging is relatively a \textit{vague} task -- The exact mechanism of annotating tags is not fully known. This aspect suits well for the fundamental motivation of deep learning, which is, to reduce inductive bias and let the data speak~\cite{nam2018deep}.   


The goal of vocal melody extraction is to estimate the F0 frequency of the (dominant) vocal track in given mixtures. Various deep learning methods have been adopted: a fully-connected neural network with Hidden Markov Model \cite{kum2016melody}, a bidirectional long short-term memory network \cite{rigaud2016singing}, a CNN~\cite{bittner2017deep}, encoder-decoder networks \cite{lu2018vocal, hsieh2019streamlined} and a CRNN \cite{kum2019joint}. Recently, a frequency-temporal attention module was introduced in \cite{yu2021frequency} to learn the relevant regions for predictions. Some special representations are proposed including HCQT~\cite{bittner2017deep}, a combination of frequency and periodicity~\cite{su2015combining}, and source-separated tracks \cite{jansson2019joint, gao2021vocal}.


Chord recognition is a MIR task to ``produce a time-varying symbolic representation of the signal in terms of chord labels'' \cite{mcfee2017structured}. Compared to music tagging, we clearly understand how chords of music signals can be decided -- They are based on the combination of the present musical notes. Therefore, models have been designed to take advantage of note representations such as constant-Q transform (CQT) or chromagram.
The early deep learning-based chord recognition models are based on a RNN \cite{boulanger2013audio} and a CNN \cite{humphrey2012rethinking}. Later, a CRNN has been used in~\cite{mcfee2017structured} to combine the merits of RNNs and CNNs.
More recently, (bi-directional) Transformer was used, achieving state-of-the-art performance~ \cite{park2019bi, chen2019harmony}. 




\section{Methods}\label{sec:method}

\begin{figure}
 \centerline{
 \includegraphics[width=\columnwidth]{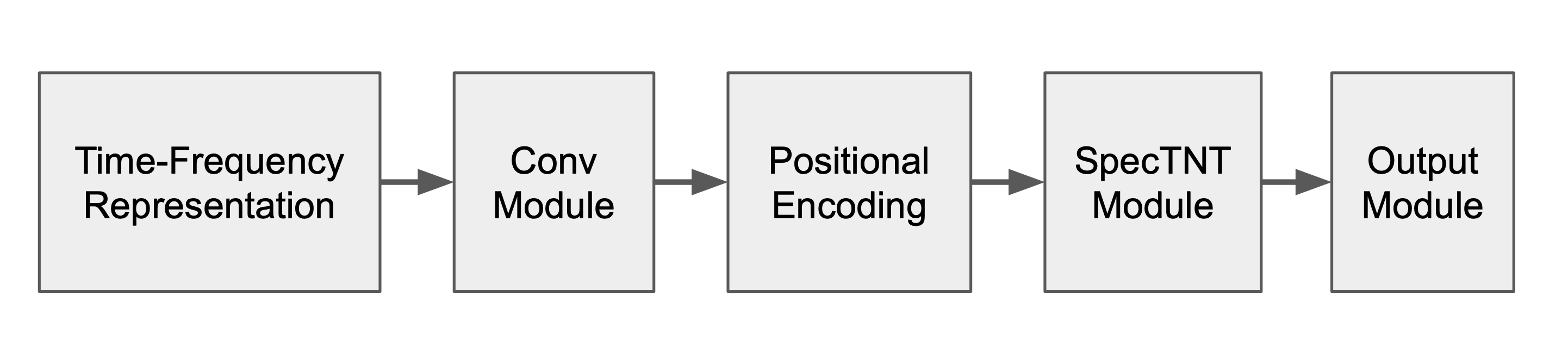}}
 \caption{The block diagram of the whole SpecTNT. The details of positional encoding and SpecTNT module are illustrated in Figure~\ref{fig: position} and \ref{fig:tnt}, respectively.}
 \label{fig:spectnt}
\end{figure}

As illustrated in Figure \ref{fig:spectnt}, the proposed SpecTNT architecture consists of a convolutional module, positional encoding, SpecTNT module, and output module. 

The input time-frequency representation is first processed with a stack of convolutional layers for local feature aggregation. Then, the positional information is added to the data. In the SpecTNT module, the intermediate representation is fed into a stack of SpecTNT blocks. 
Lastly, the output module projects the final embedding into the desired dimension for different tasks. We detail each module in the following subsections. 

\subsection{Convolutional module}

The purpose of this convolutional module is to employ different strategies for generating intermediate representations with pooling or striding convolution techniques depending on the nature of the task.
Let the input time-frequency representation be $S \in \mathbb{R}^{T \times F \times K}$ where $T$ is the number of time-steps, $F$ is the number of frequency bins, and $K$ is the number of channels.
$S$ is first passed into a stack of convolutional layers. 
We utilize the residual unit proposed in \cite{he2016identity} to be the basic building block of the convolutional module.
The representation after the convolutional module is denoted as $S' = [S'_1, S'_2, ..., S'_{\hat{T}}] \in \mathbb{R}^{\hat{T} \times \hat{F} \times \hat{K}}$, where $\hat{F}$, $\hat{T}$, and $\hat{K}$ are the numbers of frequency bins, time-steps, and channels, respectively.


\begin{figure}
 \centerline{
 \includegraphics[width=1.0\columnwidth]{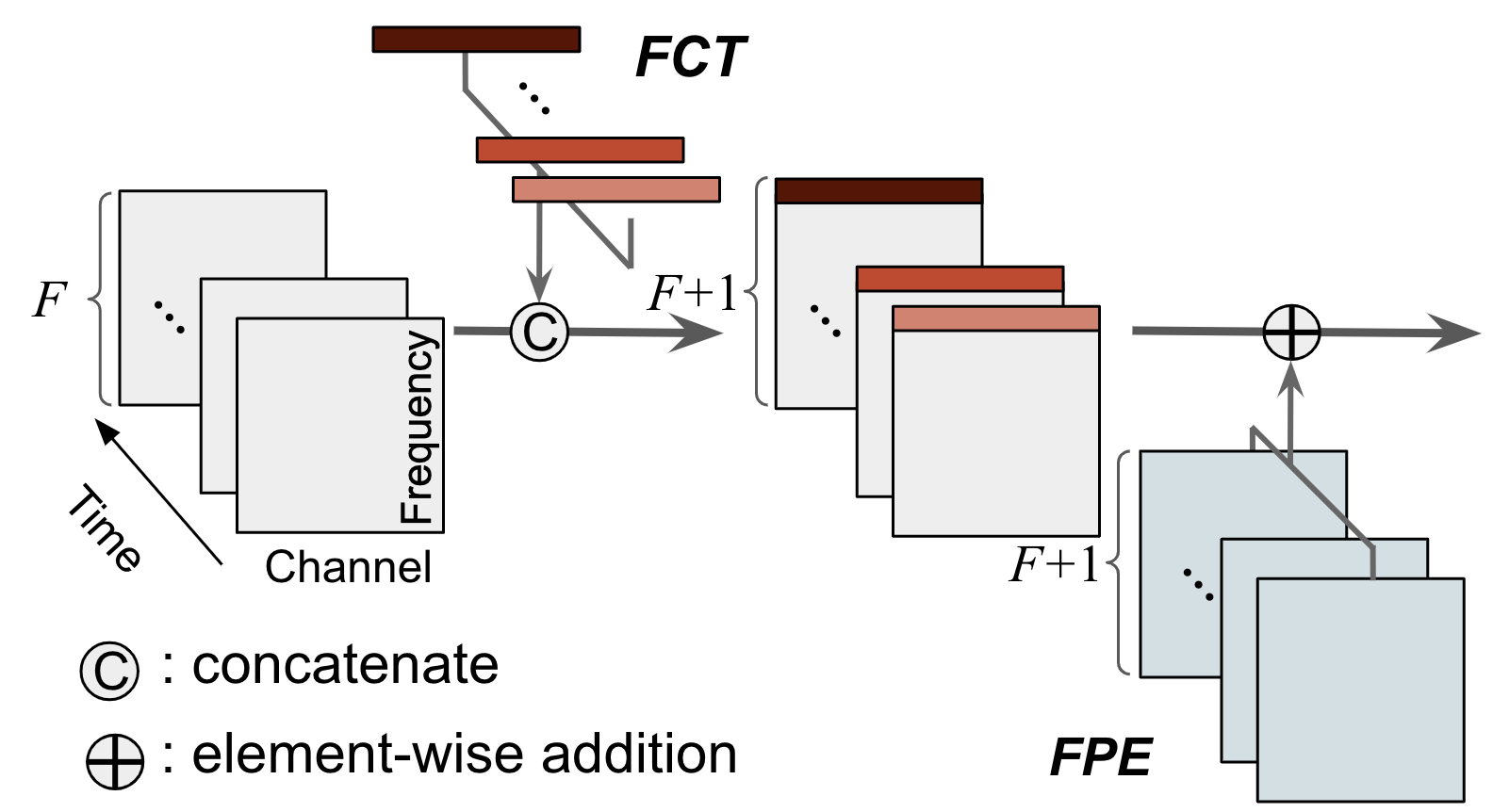}}
 \caption{An illustration of the application of Frequency class token (FCT) and frequency positional encoding (FPE). $F$ refers to the number of frequency bins of the input time-frequency representation.}
 \label{fig: position}
\end{figure}

\subsection{Frequency Class Token}

As depicted in Figure \ref{fig: position}, \emph{Frequency class token} (FCT) is an embedding vector initialized with all zeros to serve as the placeholder and defined as $c_t = \mathbf{0}^{1 \times \hat{K}}$. 
Let $S'_t \in \mathbb{R}^{\hat{F} \times \hat{K}}$ denote 
the input data at each time-step $t$.
The input data and FCT are concatenated as following:
\begin{equation} \label{eq: concat_fct}
S''_t = \mathrm{Concat}[c_t, S'_t]. 
\end{equation}
Here, the role of FCT $c_t$ is similar to the classification token~\cite{devlin2018bert}. 
It is expected to extract spectral features from each frequency bin of the $t$-th frame during the spectral self-attention in the later stages.

\subsection{Positional encoding}

In the original Transformer paper, a sinusoidal positional encoding was added to the input sequence to make the following layers aware of the order of input elements~\cite{vaswani2017attention}.
From a similar motivation, we adopt a learnable positional embedding to encode the sequence order of frequency bins. 

We encode the positional information of frequencies by adding the frequency positional embedding (FPE) to the data $S''$. FPE is a learnable matrix $E^\phi \in \mathbb{R}^{(\hat{F}+1) \times \hat{K}}$. The addition process is done at each time-step $t$:  
\begin{equation} \label{eq: fpe_emb}
    \hat{S}_t = S''_t \oplus E^\phi,
\end{equation}
where $\oplus$ is the element-wise addition, and the resulting FCTs are denoted by $\hat{C} = [\hat{c}_1, \hat{c}_2, ... , \hat{c}_{\hat{T}}]$.
Then, the resulting representation $\hat{S}_t$ is able to carry information about pitch and timbre to the following attention layers. For example, a pitch in the signal can lead to high energy at a specific frequency bin, and the positional embedding makes FCT aware of the position of that frequency 


\begin{figure}
 \centerline{
 \includegraphics[width=\columnwidth]{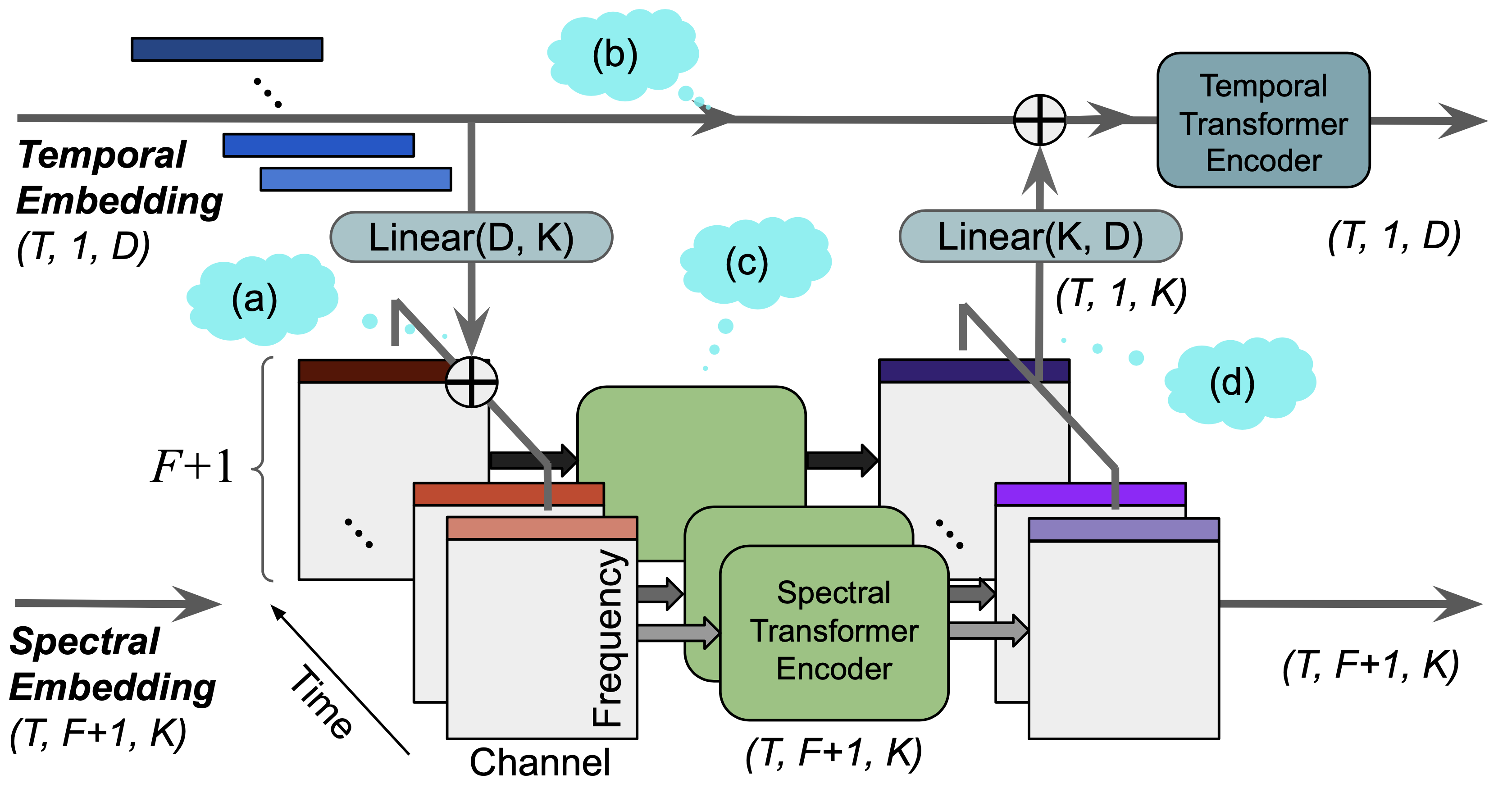}}
 \caption{The block diagram of a SpecTNT block. Tensors and modules are illustrated with non-rounded and rounded rectangles, respectively. We specify the non-batch shape of tensors for clarity, and explain (a) -- (d) in the main text.
}
 \label{fig:tnt}
\end{figure}

\subsection{Transformer in Transformer (TNT)}


Inspired by the architecture in \cite{han2021transformer}, 
we design a SpecTNT block to handle audio data, as depicted in Figure~\ref{fig:tnt}. The SpecTNT block holds two data flows: \emph{spectral embedding} (SE) and \emph{temporal embedding} (TE). The two data flows are respectively processed with two Transformer encoders, namely \emph{temporal Transformer} and \emph{spectral Transformer}.
Because the SpecTNT block is repeated multiple times in the SpecTNT module, we introduce a notation $l$ to specify the layer index for both SE and TE.

In the following sections, we explain each component of a SpecTNT block (Section \ref{sec:TE} through Section \ref{sec: transformer_enc}) and the entire procedure (Section \ref{sec: TNT}).

\subsubsection{Temporal Embedding} \label{sec:TE}

In the proposed model, we introduce the temporal embedding (TE) to distribute the information of FCTs across the time axis. We can write the TE at layer $l$ as:
\begin{equation} \label{eq: te}
E^l = [e^l_1, e^l_2, ... , e^l_{\hat{T}}],
\end{equation}
where $e^l_t \in \mathbb{R}^{1 \times D}$ is a
TE vector at time $t$ and $D$ is the number of features. In practice, TE is a learnable matrix and is initialized randomly as $E^0 \in \mathbb{R}^{\hat{T} \times D}$ prior to entering the first SpecTNT block.
 
There are two bridges between the spectral and temporal data flows. 
We use FCTs, the first frequency bin of SEs, for this communication. First, TE sends information to FCTs by passing $e^l_t$ to a linear projection layer. Then, the projected $D$-dimensional vectors are added to FCTs (Figure~\ref{fig:tnt}-(a)). Second, after spectral transformer encoder (Figure~\ref{fig:tnt}-(c)), FCTs (purple arrays) are projected back to $K$-dimension (Figure~\ref{fig:tnt}-(d)). Note that TE also has a skip-connection (Figure~\ref{fig:tnt}-(b)).  
 

\subsubsection{Spectral Embedding}

The output from the positional encoding, $\hat{\mathcal{S}}$, will serve as an input SE for the first SpecTNT block and is denoted as $\hat{\mathcal{S}}^0$. As mentioned above, SE includes FCTs, which help aggregate useful spectral information from the local. As a general notation, we write the data flow of SE as:
\begin{equation} \label{eq: se_emb}
    \hat{\mathcal{S}}^l = \left[[\hat{c}^l_1, \hat{S}^l_1], [\hat{c}^l_2, \hat{S}^l_2],..., [\hat{c}^l_{\hat{T}}, \hat{S}^l_{\hat{T}}] \right],
\end{equation}
where $l=0,1,...,L$, and $c^l_t$ and $\hat{S}^l_t$ are respectively the FCTs of $l$-th layer and spectral data at time-step $t$.
Then, SE can interact with the TE through FCTs, so the local spectral features can be processed in a temporal and global manner.

\subsubsection{Transformer Encoder} \label{sec: transformer_enc}

A Transformer encoder is composed of three components: multi-head self-attention (MHSA), feed-forward network (FFN), and layer normalization (LN). 

Self-attention (SA) \cite{vaswani2017attention} plays the pivotal role in a Transformer encoder. It takes three inputs: $Q \in \mathbb{R}^{T \times d_q}$, $K \in \mathbb{R}^{T \times d_k}$ and $V \in \mathbb{R}^{T \times d_v}$ which represent the queries, keys, and values, respectively. $T$ is the number of time-steps, and $d_q$, $d_k$ and $d_v$ indicate the dimension of features for $Q$, $K$, and $V$, respectively. The output is the weighted sum over the values based on the dot product similarity between queries and keys at the corresponding time-step.

The $\mathrm{MHSA}$ module \cite{vaswani2017attention} is an extension of SA. It splits the three inputs $Q$, $K$ and $V$ along their feature dimension into $h$ number of ``heads'' and performs multiple SA's, each on a head, in parallel. The outputs of heads are then concatenated and linearly projected into the final output.
The $\mathrm{FFN}$ module has two linear layers with a GELU activation function in the middle. We also adopt the pre-norm residual units \cite{xiong2020layer} to stabilize the training.

With the three components, the Transformer encoder (either spectral or temporal) is built and denoted by 
\begin{equation}
X_l = \mathrm{Enc}(X_{l-1}),
\end{equation}
where the operations within it can be written as 
\begin{equation} \label{eq: transformer_encoder}
    \begin{aligned} 
        X'_{l-1} &= X_{l-1} +  \mathrm{MHSA}(\mathrm{LN}(X_{l-1})), \\
        X_l &= X'_{l-1} + \mathrm{FFN}(\mathrm{LN}(X'_{l-1})).
    \end{aligned}
\end{equation}


\subsubsection{Stacking SpecTNT Blocks} \label{sec: TNT}

We stack three SpecTNT blocks for the SpecTNT module. The module starts with inputting the initial SE, $\hat{\mathcal{S}}^0$, and the initial TE, $E^0$, to the first SpecTNT block. 

For a SpecTNT block, there are four steps. 
First, each FCT vector in $\hat{\mathcal{S}}^{l-1}$ is updated by adding the linear projection of the associated TE vector (Figure~\ref{fig:tnt}-(a)):
\begin{equation} \label{eq: te2fct}
    \tilde{c}^{l-1}_t = \hat{c}^{l-1}_t \oplus \mathrm{Linear}(e^{l-1}_t),
\end{equation}
where $\mathrm{Linear}(\cdot)$ is a shared linear layer.
Second, the SE $\tilde{\mathcal{S}}^{l-1}$ (with the updated FCTs $[\tilde{c}^{l-1}_t]^{\hat{T}}_{t=1}$ at the first row) is passed through the spectral Transformer (Figure~\ref{fig:tnt}-(c)):
\begin{equation}
\hat{\mathcal{S}}^{l} = \mathrm{SpecEnc}(\tilde{\mathcal{S}}^{l-1}).
\end{equation}
Third, each FCT vector in $\hat{\mathcal{S}}^{l}$ is linearly projected and added back to the corresponding TE vector (Figure~\ref{fig:tnt}-(d)):
\begin{equation} \label{eq: fct2te}
    \tilde{e}^{l-1}_t = e^{l-1}_t \oplus \mathrm{Linear}(\hat{c}^l_t).
\end{equation}
Finally, we propose to encode only the updated TE (i.e., $\tilde{E}^{l-1}=[\tilde{e}^{l-1}_t]^{\hat{T}}_{t=1}$), instead of TE + SE, with the temporal Transformer:
\begin{equation}
E^{l} = \mathrm{TempEnc}(\tilde{E}^{l-1}).
\end{equation}
This operation builds up the relationship along the time axis and is the key role that leads to better model and data efficiency. We consider the temporal Transformer only needs to see the information of the frequency bins which are attended by the FCT and such design largely reduces the size of the model and also improves the performance on smaller datasets in preliminary experiments.






\subsection{Output Module} \label{sec:output_module}
The output TE of the 3rd SpecTNT block, $E^3$, can be used towards the final output. For frame-wise prediction tasks such as vocal melody extraction and chord recognition, we feed each TE vector $e^3_t$ into a shared fully-connected layer with sigmoid or softmax function for final output.
For song-level prediction tasks such as music tagging, we initialize a temporal class token $\epsilon^l$ ($l=0$) concatenated at the front of $E^l$:
\begin{equation} \label{eq: te_cls}
    \hat{E}^l = [\epsilon^l, e^l_1, e^l_2, ... , e^l_{\hat{T}}],
\end{equation}
Note that $\epsilon^l$ does not have an associated FCT in SE, but is for aggregating TE vectors along the time axis. Finally, we feed $\epsilon^3$ to a fully-connected layer, followed by a sigmoid layer, to get the probability output.

\section{Experiments}\label{sec:typeset_text}

In this section, we evaluate SpecTNT on various types of MIR tasks to demonstrate its effectiveness and versatility. We choose three MIR tasks -- music tagging, vocal melody extraction, and chord recognition. 

\subsection{Implementation}

\begin{table}
 \begin{center}
 \scalebox{0.8}{%
 \begin{tabular}{|l|l|l|l|l|}
  \hline
  Task & ($p_f$, $p_t$) &($k$, $d$) & ($h_k$, $h_d$) & $o_d$ \\
  \hline
  Music tagging & (1, 4) &(96, 96)  &  (4, 8) & 50\\
  Vocal melody extraction & (4, 1)& (128, 128) & (8, 8)& ($T$, 481) \\
  Chord recognition & (1, 1) &(64, 256) & (4, 8) & ($T$, 25)\\
  \hline
 \end{tabular}
 }
\end{center}
 \caption{Settings of SpecTNT for different tasks, where $p_f$ and $p_t$ represent the pooling ratio we apply along the frequency and time axis in the convolutional module, $k$ and $d$ are the feature dimension, $h_k$ and $h_d$ denotes the number of heads for the spectral and temporal transformer encoder respectively, and finally, $o_d$ represents the output dimension, $T$ indicates frame-wise predictions. }
 \label{tab:implementaiton_detail}
\end{table}

SpecTNT is implemented using Pytorch \cite{NEURIPS2019_9015}. Due to the difference in dataset sizes and the natures of tasks, we use different hyper-parameters for the tasks as shown in Table~\ref{tab:implementaiton_detail}. All models include dropout with a rate of $0.15$ in the Transformers of the TNT modules. We use AdamW \cite{loshchilov2017decoupled} as the learning optimizer. The initial learning rates are set to $10^{-3}$ for vocal melody extraction, $5 \times 10^{-4}$ for music tagging and chord recognition, and a weight decay of $5 \times 10^{-3}$ is set for all the tasks.

For the input representation of music tagging, we re-sample the audio at the 22,050~Hz and use an input length of 4.54~second. Log-magnitude mel-spectrograms are computed with 128 mel filter banks, 1024 samples of Hann window, and a hop size of 512 samples. For vocal melody extraction, input waveforms are re-sampled at the 16,000~Hz sample rate. We take 3-second segments input and their log-magnitude spectrograms are computed with 2048 samples of Hann window and a hop size of 320 samples. For chord recognition, we try two types of input representation. The first input type is 24-dimensional chroma features with a frame rate of 46~ms \cite{burgoyne2011expert}. Out of the whole track, we use 400 frames as an input. 
The second input type is CQT, which is computed from a 18.2 second audio at the 22,050~Hz sample rate. The CQT includes six octaves starting from C1 (32.70~Hz) with 24 bins per octave, and is based on a hop size of 2048.


\subsection{Ablations Study}
To validate the design choices we make, we consider three various models by progressively removing the components of SpecTNT as follows.

\textbf{A1}: Remove the operation of (a) in Figure \ref{fig:spectnt} (i.e., Eq. \ref{eq: te2fct}) and initialize the FCTs as learnable vectors.

    
\textbf{A2}: Neglect the FCTs but use the full spectral embeddings for operations (a) and (c) in Figure \ref{fig:spectnt} (i.e., Eq. \ref{eq: te2fct} and Eq. \ref{eq: fct2te}). The resulting model can be seen as using the original TNT block \cite{han2021transformer}.

\textbf{A3}: Remove the data flow of spectral embedding, so the model is reduced to the original Transformer \cite{vaswani2017attention} for aggregating the input sequence in a traditional way.

In the following evaluations of different tasks, we will include the results of the three variants for comparison. 

\subsection{Music Auto-tagging}\label{subsec:tagging}

\textbf{Datasets}~~
Million song dataset (MSD) \cite{bertin2011million} consists of one million audio previews and a subset of it has crowd-sourced music tags. Typically, a subset with the 50 most frequent tags are used with randomly split train/validation/test sets~\cite{choi2016automatic}. However, these tags are noisy and the random split without considering artist overlaps may cause unintended information leakage. Therefore, we take advantage of manually cleaned 50 tags from a previous work~\cite{won2020multimodal} and split the dataset based on artist names so that there is no overlapped artists among the training/validation/test sets. As a result, we use 233,147 tracks, of which 70\%, 15\%, and 15\% are allocated for training, validation, and test sets, respectively. During training, we apply random data augmentation to the input waveform following the pipeline introduced in \cite{spijkervet2021contrastive}.

\vspace*{0.3cm}
\noindent
\textbf{Baseline Models}~~
Two baselines methods are compared. The first is CNNSA \cite{won2019toward}, which employs a convolutional front-end and a transformer encoder to aggregate the temporal feature. The second baseline \cite{won2020evaluation} uses 7-layer short-chunk CNN with residual connection, followed by a fully-connect layer for final output. This model has shown state-of-the-art performance in music auto-tagging.
We utilize the original implementation of \cite{won2020evaluation} to train the baseline under the same configuration as our proposed model.

\vspace*{0.3cm}
\noindent
\textbf{Evaluation Metrics}~~
Area Under Precision Recall Curve (PR-AUC) and Area Under Receiver Operating Characteristic curve (ROC-AUC) are used.

\vspace*{0.3cm}
\noindent
\textbf{Results}~~
The results of music auto-tagging are summarized in Table \ref{tab:results_music_tagging}. SpecTNT outperforms prior state-of-the-art models in both metrics. In the ablation study, A1 performs the worst, while A2 and A3 show similar results to SpecTNT. This can be explained from the perspective of data distribution: the top 50 tags of MSD dataset are mostly related to genre and style, both of which need enough temporal information to characterize. In A1, the process of updating FCTs with TEs is removed and this may interfere the temporal information flow being shared with the spectral data and cause the performance drop. By looking into the precision scores of individual tags where SpecTNT outperforms A3, we observe that instrumental tags such as ``piano'' and ``guitar'' can benefit from SpecTNT, because they may require more spectral information to model well. This shows the benefit of adding the spectral transformer. Also, the smaller performance difference among SpecTNT, A2, and A3 indicates that the size of MSD dataset might be enough to support architectures with less prior knowledge. That is, A2 and A3 are able to sufficiently learn from MSD the useful information without further utilizing FCTs to interact with the temporal embeddings.



\begin{table}
 \begin{center}
 \scalebox{1}{%
 \begin{tabular}{|l||l l|}
  \hline
  Method & ROC-AUC & PR-AUC  \\
  \hline
  Short-chunk CNN + Res  & 91.55 & 37.08 \\ 
  CNNSA  & 91.57 & 37.09  \\
  \hline
  SpecTNT  & \textbf{92.08} & \textbf{38.62}  \\
  \hline
  A1 & 91.92 & 37.85 \\
  A2 & 92.07 & 38.59 \\
  A3 & 92.06 & 38.46 \\
  \hline
 \end{tabular}
 }
\end{center}
 \caption{Results (in $\%$) for automatic music tagging.}
 \label{tab:results_music_tagging}
\end{table}


\subsection{Vocal Melody Extraction}

\textbf{Datasets}~~
We use two datasets to train the models: MIR1K \cite{hsu2009improvement}, which includes 1000 Chinese karaoke clips, and a 48-song subset of MedleyDB \cite{bittner2014medleydb} that includes vocal tracks. Since the training sets are relative small, we adopt a pipeline with four steps of augmentation techniques. There is a chance for each step to be applied to a training sample:
\textit{i}) pitch-shifting by up to $\pm 2$ semitones (with 100\% chance), 
\textit{ii}) replacing the original background track with a randomly selected, different background track (with 50\% chance), 
\textit{iii}) changing the gain of the vocal within $[-4, 2]$ dB (with 100 \% chance), and 
\textit{iv}) completely removing the background track (with 10\% chance).

We choose three test sets for evaluation: ADC2004, MIREX05, and MedleyDB. For ADC2004 and MIREX05, we only use the samples that have melody sung by human voice. This results in 12 samples from ADC2004 and 9 samples from MIREX05. For MedleyDB, we only use the songs that have singing voice included in their ``MELODY2'' annotations, yielding 12 songs. The ground-truth pitches are obtained from the MELODY2 annotations within the intervals marked as ``female singer'' or ``male singer.'' These 12 songs are not included in training.

\vspace*{0.3cm}
\noindent
\textbf{Baseline Models}~~
We compare our model with two baseline models. 
The first baseline is the joint detection and classification model (JDC)~\cite{kum2019joint} based on CRNN. We use the most representative architecture, called ``Main'' in ~\cite{kum2019joint}. The second baseline is the frequency-temporal attention network (FTANet)~\cite{yu2021frequency}, which is a CNN-based model that employs attention mechanism along the frequency and time axis. 
We re-implemented JDC and FTANet using Pytorch and used the suggested hyper-parameters in \cite{kum2019joint,yu2021frequency}. Both models are trained under the same configuration (e.g., data split and augmentation process) as our model.

\vspace*{0.3cm}
\noindent
\textbf{Evaluation Metrics}~~
Overall Accuracy (OA), Raw Pitch Accuracy (RPA), and Voice Recall (VR) are adopted for evaluation. We use mir\_eval library \cite{raffel2014mir_eval} to compute the performance values with a tolerance range of 50 cents.

\vspace*{0.3cm}
\noindent
\textbf{Results}~~
Table \ref{tab: results_vocal} shows the results for vocal melody extraction. SpecTNT outperforms the baselines by a large margin in terms of RPA and VR. To the best of our knowledge, this is the first attempt to apply Transformers to this task and the results demonstrate its superiority over the CNN and CRNN counterparts. It is worth noting that FTANet is trained with an input representation specifically designed for pitch detection \cite{su2015combining}, but our model works well with spectrogram input. In addition, A3 shows the largest performance drop, and this demonstrates the usefulness of spectral Transformer when training on smaller data. 


\begin{table}
 \footnotesize  
 \begin{tabular}{|p{1cm}||p{.32cm}p{.32cm}p{.34cm}|p{.34cm}p{.34cm}p{.35cm}|p{.32cm}p{.32cm}p{.35cm}|}
\hline
 Dataset & \multicolumn{3}{c|}{ADC2004}  &  \multicolumn{3}{c|}{MIREX05}  & \multicolumn{3}{c|}{MedelyDB}  \\
\hline
Method & OA & RPA & VR & OA & RPA & VR & OA & RPA & VR \\
\hline
JDC  & 71.2 & 68.1 & 73.1 & 86.0 & 80.7 & 85.8 & 77.0 & 64.8 & 73.9 \\ 
FTANet & 71.2 & 69.3 & 72.9 & 89.9 & 86.5 & 91.2 & \textbf{79.4} & 66.0 & 72.0\\
\hline
SpecTNT  & \textbf{85.3} & \textbf{85.0} & 88.3 & \textbf{91.7} & \textbf{90.4} & \textbf{95.2} & 78.4 & \textbf{77.9} & \textbf{87.4} \\
\hline
A1  & 84.8 & 84.2 & \textbf{89.1} & 90.2 & 88.3 & 94.1 & 77.9 & 75.0 & 85.4 \\
A2  & 84.9 & 84.5 & 88.3          & 89.7 & 87.7 & 93.1 & 79.3 & 75.6 & 84.0 \\
A3  & 84.5 & 83.7 & 87.2          & 88.9 & 87.2 & 92.0 & 74.5 & 72.7 & 83.1 \\
\hline
\end{tabular}
\caption{Results (in $\%$) for vocal melody extraction.}
\label{tab: results_vocal}
\end{table}

\subsection{Chord Recognition}

\vspace*{0.3cm}
\textbf{Datasets}~~
We use the Billboard dataset to evaluate SpecTNT for the chord recognition task. The dataset contains 890 pieces selected from the Billboard chart slots \cite{burgoyne2011expert}. 
Following \cite{chen2019harmony}, duplicates pieces are first removed to leave 739 unique pieces in total. The official release of the dataset only comes with 24-D chroma vectors, which might be insufficient to fully demonstrate the effectiveness of SpecTNT. Therefore, we manually collected the audio files based on the provided meta-data. Due to the potential version mismatch between our audio files and that for official chord annotations, we applied dynamic time warping (DTW) \cite{muller2007dynamic} to validate each song. Specifically, we first replicated the chroma features of the official release using Sonic Annotator \cite{chris2010a} on our audio files, and then calculated the alignment cost between the two versions of chromagrams for each song using DTW. We selected 462 songs with the lowest alignment costs.
The songs with ID's smaller than 1000 are used for training and the remaining for testing. 
To augment the training data (chroma and audio), we shifted the pitches by up to $\pm 6$ semitones.
For evaluation, we adopt the ``maj/min'' label set with 25 classes, where 24 are major and minor triads across the 12 semitones plus an additional ``no chord'' class.

\vspace*{0.3cm}
\noindent
\textbf{Baseline Models}~~
We compare to three baseline models: \textit{i}) CR2 model from ~\cite{mcfee2017structured}, which is a CRNN-based model, \textit{ii}) a bi-directional Transformer (BTC) \cite{park2019bi}, and \textit{iii}) Harmony Transformer (HT) \cite{chen2019harmony}. BTC and HT are known to be the current state-of-the-art models for chord recognition. For CR2 and BTC, we use the official implementations with the suggested default settings for both chromagram and CQT inputs. For HT, we report the chromagram-based results in \cite{chen2019harmony}, since the train/test data split and data augmentation are very similar to us. We did not conduct experiments using HT with CQT input because non-trivial modifications are required for the model.

\vspace*{0.3cm}
\noindent
\textbf{Evaluation Metrics}~~
The Weighted Chord Symbol Recall (WCSR) score is reported as evaluation metric. WCSR is the percentage of correctly identified frames and can be computed by $\frac{t_c}{t_a} \times 100 (\%)$,
where $t_c$ is the duration of the correctly predicted segments, and $t_a$ is the total duration of the test segments. 

\begin{table}
 \begin{center}
 \scalebox{1}{%
 \begin{tabular}{|l||l|l|}
  \hline
  Method & Chroma & CQT\\
  \hline
  CR2 & 78.92 & 73.38 \\ 
  BTC & 77.98 & 73.92 \\
  HT & \textbf{82.68} & - \\
  \hline
  SpecTNT & 80.47 & \textbf{75.62} \\
  \hline
  A1 & 80.10 & 74.83 \\
  A2 & 78.76 & 74.44\\
  A3 & 77.69 & 74.99\\
  \hline
 \end{tabular}
 }
\end{center}
 \caption{Results (in $\%$) for chord recognition task}
 \label{tab:Chord_results}
\end{table}

\vspace*{0.3cm}
\noindent
\textbf{Results}~~
Table \ref{tab:Chord_results} shows the results for chord recognition. For ``Chroma'' case, the full Billboard dataset is used. For ``CQT'' case, the 462 songs with audio are used. From the results, SpecTNT can outperform all the baselines except HT (with chromagram input). However, HT may benefit from joint training with an additional segmentation loss, so the comparison could be unfair. Compared to BTC and CR2, SpecTNT achieves better performance for both types of input. For the ablation study, since we used less data for CQT input, A2, which is the largest model, may suffer from over-fitting and thus performs the worst.





 

\section{Conclusion}
We proposed SpecTNT, a novel Transformer architecture that models spectrograms along both the time and frequency axes. 
The introduction of FCT enables effective communication between the spectral embeddings and temporal embeddings, maximizing the benefit of Transformer encoder for flexible, local, and global modeling.
In experiments, SpecTNT has demonstrated state-of-the-art performance in music tagging and vocal melody extraction and shown competitive performance in chord recognition. 
For future work, we plan to apply SpecTNT to other MIR tasks, such as beat tracking and structure segmentation. 

\bibliography{ISMIR2021_template}

\end{document}